\documentclass[12pt]{article}

\usepackage{amsmath,amsfonts,enumerate,array}
\usepackage{graphicx}
\usepackage[usenames]{color}
\usepackage[english]{babel}
\usepackage{fancybox}
\usepackage{chngpage} 
\usepackage{amssymb}
\usepackage{float}

\newcommand{\captionfonts}{\small}

\makeatletter  
\long\def\@makecaption#1#2{%
  \vskip\abovecaptionskip
  \sbox\@tempboxa{{\captionfonts #1: #2}}%
 \ifdim \wd\@tempboxa >\hsize
    {\captionfonts #1: #2\par}
  \else
    \hbox to\hsize{\hfil\box\@tempboxa\hfil}%
  \fi
  \vskip\belowcaptionskip}
\makeatother   

\usepackage{mciteplus}

\usepackage[colorlinks=true,      linkcolor=blue,      urlcolor=blue,      filecolor=blue,      citecolor=blue,
      pdfstartview=FitH,     pdfpagemode=UseNone,      bookmarksopen=true]{hyperref}  
      
\usepackage[all]{hypcap}     

\begin{document}

\numberwithin{equation}{section}


\mathchardef\mhyphen="2D


\makeatletter
\newcommand{\contraction}[5][1ex]{%
  \mathchoice
    {\contraction@\displaystyle{#2}{#3}{#4}{#5}{#1}}%
    {\contraction@\textstyle{#2}{#3}{#4}{#5}{#1}}%
    {\contraction@\scriptstyle{#2}{#3}{#4}{#5}{#1}}%
    {\contraction@\scriptscriptstyle{#2}{#3}{#4}{#5}{#1}}}%
\newcommand{\contraction@}[6]{%
  \setbox0=\hbox{$#1#2$}%
  \setbox2=\hbox{$#1#3$}%
  \setbox4=\hbox{$#1#4$}%
  \setbox6=\hbox{$#1#5$}%
  \dimen0=\wd2%
  \advance\dimen0 by \wd6%
  \divide\dimen0 by 2%
  \advance\dimen0 by \wd4%
  \vbox{%
    \hbox to 0pt{%
      \kern \wd0%
      \kern 0.5\wd2%
      \contraction@@{\dimen0}{#6}%
      \hss}%
    \vskip 0.2ex%
    \vskip\ht2}}
\newcommand{\contracted}[5][1ex]{%
  \contraction[#1]{#2}{#3}{#4}{#5}\ensuremath{#2#3#4#5}}
\newcommand{\contraction@@}[3][0.06em]{%
  \hbox{%
    \vrule width #1 height 0pt depth #3%
    \vrule width #2 height 0pt depth #1%
    \vrule width #1 height 0pt depth #3%
    \relax}}
\makeatother

\newcommand{\be}{\begin{equation}} 
\newcommand{\ee}{\end{equation}} 
\newcommand{\bea}{\begin{eqnarray}\displaystyle}
\newcommand{\eea}{\end{eqnarray}}
\newcommand{\bt}{\begin{tabular}}
\newcommand{\et}{\end{tabular}}
\newcommand{\bs}{\begin{split}}
\newcommand{\es}{\end{split}}
\def\r{\rightarrow}

\newcommand{\nsnsket}{|0_{NS}\rangle^{(1)}\newotimes |0_{NS}\rangle^{(2)}}					
\newcommand{\nsnsbra}{{}^{(1)}\langle 0_{NS}| \newotimes {}^{(2)}\langle 0_{NS}|}		
\newcommand{\nsket}{|0_{NS}\rangle}																									
\newcommand{\nsbra}{\langle 0_{NS}|}																								
\newcommand{\nstket}{|0_{NS}\rangle_t}																										
\newcommand{\nstbra}{{}_t\langle 0_{NS}|}																								
\newcommand{\rmmket}{|0_R^-\rangle^{(1)}\newotimes |0_R^-\rangle^{(2)}}							
\newcommand{\rmmbra}{{}^{(1)}\langle 0_{R,-}| \newotimes {}^{(2)}\langle 0_{R,-}|}	
\newcommand{\rmket}{|0_R^-\rangle}																									
\newcommand{\rmbra}{\langle 0_{R,-}|}																								
\newcommand{\rmtket}{|0_R^-\rangle_t}																								
\newcommand{\rmtbra}{{}_t\langle 0_{R,-}|}																					
\newcommand{\rppket}{|0_R^+\rangle^{(1)}\newotimes |0_R^+\rangle^{(2)}}							
\newcommand{\rppbra}{{}^{(1)}\langle 0_{R,+}| \newotimes {}^{(2)}\langle 0_{R,+}|}	
\newcommand{\rpket}{|0_R^+\rangle}																									
\newcommand{\rpbra}{\langle 0_{R,+}|}																								
\newcommand{\rptket}{|0_R^+\rangle_t}																								
\newcommand{\rptbra}{{}_t\langle 0_{R,+}|}																					
\newcommand{\rpmket}{| 0_R^+\rangle^{(1)} \newotimes | 0_R^-\rangle^{(2)}}					
\newcommand{\rpmbra}{{}^{(1)}\langle 0_{R,+}| \newotimes {}^{(2)}\langle 0_{R,-}|}	
\newcommand{\rmpket}{| 0_R^-\rangle^{(1)} \newotimes | 0_R^+\rangle^{(2)}}					
\newcommand{\rmpbra}{{}^{(1)}\langle 0_{R,-}| \newotimes {}^{(2)}\langle 0_{R,+}|}	

\newcommand{\nsutvket}{|0_{NS}\rangle^{(1)}\otimes |0_{NS}\rangle^{(2)}}
\newcommand{\nsutvbra}{{}^{(1)}\langle 0_{NS}| \otimes {}^{(2)}\langle 0_{NS}|}
\newcommand{\nstvket}{|0_{NS}\rangle}
\newcommand{\nstclose}{0_{NS}\rangle}
\newcommand{\nstvbra}{\langle 0_{NS}|}
\newcommand{\nstpket}{|0\rangle_t}
\newcommand{\nstpbra}{{}_t\langle 0|}
\newcommand{\rmutvket}{|0_R^-\rangle^{(1)}\otimes |0_R^-\rangle^{(2)}}
\newcommand{\rmutvbra}{{}^{(1)}\langle 0_{R,-}| \otimes {}^{(2)}\langle 0_{R,-}|} 
\newcommand{\rmtvket}{|0_R^-\rangle}
\newcommand{\rmtvbra}{\langle 0_{R,-}|}
\newcommand{\rmtpket}{|0_R^-\rangle_t}
\newcommand{\rmtpbra}{{}_t\langle 0_{R,-}|}
\newcommand{\rputvket}{|0_R^+\rangle^{(1)}\otimes |0_R^+\rangle^{(2)}}
\newcommand{\rputvbra}{{}^{(1)}\langle 0_{R,+}| \otimes {}^{(2)}\langle 0_{R,+}|} 
\newcommand{\rptvket}{|0_R^+\rangle}
\newcommand{\rptvbra}{\langle 0_{R,+}|}
\newcommand{\rptpket}{|0_R^+\rangle_t}
\newcommand{\rptpbra}{{}_t\langle 0_{R,+}|}
\newcommand{\stp}{\sigma_2^+}
\newcommand{\stm}{\sigma_2^-}

\renewcommand{\a}{\alpha}	
\renewcommand{\b}{\beta}
\newcommand{\g}{\gamma}		
\newcommand{\G}{\Gamma}
\renewcommand{\d}{\delta}
\newcommand{\D}{\Delta}
\renewcommand{\c}{\chi}			
\newcommand{\C}{\Chi}
\newcommand{\p}{\psi}			
\renewcommand{\P}{\Psi}
\newcommand{\s}{\sigma}		
\renewcommand{\S}{\Sigma}
\renewcommand{\t}{\tau}		
\newcommand{\e}{\epsilon}
\newcommand{\n}{\nu}
\newcommand{\m}{\mu}
\renewcommand{\r}{\rho}
\renewcommand{\l}{\lambda}

\newcommand{\nn}{\nonumber\\} 		
\newcommand{\newotimes}{}  				
\newcommand{\diff}{\,\text{d}}		
\newcommand{\h}{{1\over2}}				
\newcommand{\Gf}[1]{\G \Big{(} #1 \Big{)}}	
\newcommand{\floor}[1]{\left\lfloor #1 \right\rfloor}
\newcommand{\ceil}[1]{\left\lceil #1 \right\rceil}

\def\cA{{\cal A}} \def\cB{{\cal B}} \def\cC{{\cal C}}
\def\cD{{\cal D}} \def\cE{{\cal E}} \def\cF{{\cal F}}
\def\cG{{\cal G}} \def\cH{{\cal H}} \def\cI{{\cal I}}
\def\cJ{{\cal J}} \def\cK{{\cal K}} \def\cL{{\cal L}}
\def\cM{{\cal M}} \def\cN{{\cal N}} \def\cO{{\cal O}}
\def\cP{{\cal P}} \def\cQ{{\cal Q}} \def\cR{{\cal R}}
\def\cS{{\cal S}} \def\cT{{\cal T}} \def\cU{{\cal U}}
\def\cV{{\cal V}} \def\cW{{\cal W}} \def\cX{{\cal X}}
\def\cY{{\cal Y}} \def\cZ{{\cal Z}}

\def\mC{\mathbb{C}} \def\mP{\mathbb{P}}  
\def\mR{\mathbb{R}} \def\mZ{\mathbb{Z}} 
\def\mT{\mathbb{T}} \def\mN{\mathbb{N}}
\def\mH{\mathbb{H}} \def\mX{\mathbb{X}}
\def\CP{\mathbb{CP}}
\def\RP{\mathbb{RP}}
\def\Z{\mathbb{Z}}
\def\N{\mathbb{N}}
\def\H{\mathbb{H}}

\newcommand{\Zd}{\ensuremath{ Z^{\dagger}}}
\newcommand{\Xd}{\ensuremath{ X^{\dagger}}}
\newcommand{\Ad}{\ensuremath{ A^{\dagger}}}
\newcommand{\Bd}{\ensuremath{ B^{\dagger}}}
\newcommand{\Ud}{\ensuremath{ U^{\dagger}}}
\newcommand{\Td}{\ensuremath{ T^{\dagger}}}
\newcommand{\T}[3]{\ensuremath{ #1{}^{#2}_{\phantom{#2} \! #3}}}		
\newcommand{\tr}{\operatorname{tr}}
\newcommand{\sech}{\operatorname{sech}}
\newcommand{\Spin}{\operatorname{Spin}}
\newcommand{\Sym}{\operatorname{Sym}}
\newcommand{\Com}{\operatorname{Com}}
\def\adj{\textrm{adj}}
\def\id{\textrm{id}}
\def\pb{\ov\psi}
\def\pt{\widetilde{\psi}}
\def\at{\widetilde{\a}}
\def\cb{\ov\chi}
\def\db{\bar\partial}
\def\delb{\bar\partial}
\def\dbar{\ov\partial}
\def\dag{\dagger}
\def\dalpha{{\dot\alpha}}
\def\dbeta{{\dot\beta}}
\def\dgamma{{\dot\gamma}}
\def\ddelta{{\dot\delta}}
\def\ad{{\dot\alpha}}
\def\bd{{\dot\beta}}
\def\dg{{\dot\gamma}}
\def\dd{{\dot\delta}}
\def\th{\theta}
\def\Th{\Theta}
\def\eb{{\ov \epsilon}}
\def\gb{{\ov \gamma}}
\def\wb{{\ov w}}
\def\Wb{{\ov W}}
\def\D{\Delta}
\def\DD{\Delta^\dag}
\def\Db{\ov D}
\def\ov{\overline}
\def\Slash{\, / \! \! \! \!}
\def\dslash{\partial\!\!\!/} 
\def\Dslash{D\!\!\!\!/\,\,}
\def\fslash#1{\slash\!\!\!#1}
\def\Fslash#1{\slash\!\!\!\!#1}
\def\del{\partial}
\def\delb{\bar\partial}
\newcommand{\ex}[1]{{\rm e}^{#1}} 
\def\ii{{i}}
\def\b{\bigskip}
\newcommand{\vs}[1]{\vspace{#1 mm}}
\newcommand{\ve}{{\vec{\e}}}
\newcommand{\shalf}{\frac{1}{2}}
\newcommand{\lb}{\rangle}
\newcommand{\al}{\ensuremath{\alpha'}}
\newcommand{\ap}{\ensuremath{\alpha'}}
\newcommand{\ft}[2]{{\textstyle {\frac{#1}{#2}} }}

\newcommand{\rmd}{\mathrm{d}}
\newcommand{\rmx}{\mathrm{x}}
\def\tA{ {\widetilde A} } 
\def\one{{\hbox{\kern+.5mm 1\kern-.8mm l}}}
\def\zero{{\hbox{0\kern-1.5mm 0}}}
\def\eq#1{(\ref{#1})}
\newcommand{\secn}[1]{Section~\ref{#1}}
\newcommand{\tbl}[1]{Table~\ref{#1}}
\newcommand{\fig}{Fig.~\ref}
\def\sqi{{1\over \sqrt{2}}}
\newcommand{\hsp}{\hspace{0.5cm}}
\def\half{{\textstyle{1\over2}}}
\let\ci=\cite \let\re=\ref
\let\se=\section \let\sse=\subsection \let\ssse=\subsubsection
\newcommand{\dpb}{D$p$-brane}
\newcommand{\dpbs}{D$p$-branes}
\def\gh{{\rm gh}}
\def\sgh{{\rm sgh}}
\def\NS{{\rm NS}}
\def\R{{\rm R}}
\def\Qp{Q_{\rm P}}
\def\QP{Q_{\rm P}}
\newcommand\dott[2]{#1 \! \cdot \! #2}
\def\eo{\overline{e}}
\newcommand{\bb}{\bigskip}
\newcommand{\ac}[2]{\ensuremath{\{ #1, #2 \}}}
\renewcommand{\ell}{l}
\newcommand{\z}{\ell}
\newcommand{\bm}{\bibitem}
\newcommand\com[2]{[#1,\,#2]}

\newcommand{\bra}[1]{{\langle {#1} |\,}}
\newcommand{\ket}[1]{{\,| {#1} \rangle}}
\newcommand{\braket}[2]{\ensuremath{\langle #1 | #2 \rangle}}
\newcommand{\Braket}[2]{\ensuremath{\langle\, #1 \,|\, #2 \,\rangle}}
\newcommand{\norm}[1]{\ensuremath{\left\| #1 \right\|}}
\def\corr#1{\left\langle \, #1 \, \right\rangle}
\def\vac{|0\rangle}


\vspace{16mm}

 \begin{center}
{\LARGE Thermalization in the D1D5 CFT}

\vspace{18mm}
{\bf   Shaun Hampton\footnote{hampton.197@osu.edu} and Samir D. Mathur\footnote{mathur.16@osu.edu}
\\}
\vspace{15mm}
Department of Physics,\\ The Ohio State University,\\ Columbus,
OH 43210, USA\\ 
\vspace{8mm}
\end{center}

\vspace{10mm}

\thispagestyle{empty}
\begin{abstract}

\vspace{3mm}

It is generally agreed that black hole formation in gravity corresponds to thermalization in the dual CFT. It is sometimes argued that if the CFT evolution shows evidence of large redshift in gravity, then we have seen black hole formation in the CFT. We argue that this is not the case: a clock falling towards the horizon increases its redshift but remains intact as a clock; thus it is not `thermalized'. Instead, thermalization should correspond to a new phase after the phase of large redshift, where the infalling object turns into fuzzballs on reaching within planck distance of the horizon. We compute simple examples of the scattering vertex in the D1D5 CFT which, after many iterations, would lead to thermalization. An initial state made of two left-moving and two right-moving excitations corresponds, in gravity,  to two gravitons heading towards each other. The thermalization vertex in the CFT breaks these excitations into multiple excitations on the left and right sides; we compute the amplitudes for several of these processes. We find secular terms that grow as $t^2$ instead of oscillating with $t$; we conjecture that this may be a  feature of processes leading to thermalization.

\end{abstract}
\newpage

\section{Introduction}

The AdS/CFT correspondence \cite{maldacena,gkp,witten} gives a remarkable map between gravity in AdS space and a CFT without gravity. In this correspondence, black hole formation in gravity is expected to map to thermalization in the dual CFT \cite{witten}. But when does the black hole form on the gravity side, and what exactly is thermalization in the CFT?

Consider an object falling towards the horizon of a black hole. As it gets closer and closer to the horizon, its redshift becomes larger and larger. It is sometimes said that this large redshift in gravity is the signal of black hole formation, and if we see the corresponding redshift in the CFT, then we would be seeing thermalization in the CFT \cite{redshift}. We will argue that this is not the case: redshift is different from thermalization. We will then perform a weak coupling computation in the CFT which we argue gives a signal for actual thermalization, albeit in a very simplified setting since we take parameters of the theory where the black hole is very small. 

\subsection{Redshift versus thermalization}

 Consider the Poincare patch  geometry created by a stack of D3 branes
 \be
 ds^2=(1+{Q\over r^4})^{-1} [-dt^2+dy_idy_i]+(1+{Q\over r^4})^{-1}(dr^2+d\Omega_5^2)
 \label{one}
 \ee
 where $y_i, i=1, \dots 3$ are coordinates along the D3 branes.  The directions $y_i$ are not compactified, so they describe an infinite plane. An object falling towards $r\rightarrow 0$ feels an increasing redshift, and the redshift diverges at $r\rightarrow 0$. Thus we have a situation with no black hole, but where we do have  a diverging redshift. The question is: should we call the corresponding process in the CFT a process of thermalization?
 
Let the infalling object be a string that is oscillating in some particular mode with a frequency $\omega$; here  $\omega$ remains a constant in the rest frame of the string. We can imagine these oscillations as the ticking of a `clock' as the clock falls towards $r=0$. 

Our clock ticks more and more slowly as it gets towards $r=0$ in the metric (\ref{one}), or as it gets near the horizon of a black hole in a black hole geometry. But the clock {\it is still intact as a clock, because it is still ticking at the  regular intervals}. The clock has not been `destroyed': a destroyed clock would not tick at regular intervals. 

Now consider the dual CFT. The states in the CFT are in an exact 1-1 correspondence with the states in the gravity theory; this is after all just the idea of AdS/CFT duality. The oscillating string in the gravity picture maps to a complicated set of gluons, but these gluons must be in a state that exhibits the same periodic oscillations, after we separate out the effect of the infall towards $r=0$. As long as the state has these well defined oscillations, we would argue that it has {\it not} thermalized. Thus if we find the CFT description of the phenomenon of gravitational redshift, then we have not obtained thermalization in the CFT. 

To summarize, redshift is a slowdown of evolution: it happens in gravity, and maps to a similar slowdown in the CFT \cite{kaplan}.  But redshift, however large, is a phenomenon distinct from thermalization.   Thermalization involves a randomization over accessible states with similar energies, such that the characteristic features on the initial state -- like a periodic oscillation -- get {\it obliterated}. 

\subsection{Black hole formation in the fuzzball paradigm}

Clearly, part of the difficulty we are facing in finding the dual of black hole formation is that we are not addressing what happens to the infalling object after it actually reaches the horizon radius $r=r_h$. That is, we are describing the increasing redshift as the infalling object approaches the horizon, but we are not addressing (i) how the CFT is supposed to describe the process of actually reaching the horizon $r=r_h$ or  (ii) if there is a CFT analogue of the classically predicted behavior of smoothly passing through the horizon into an interior region $r<r_h$. 

In the fuzzball paradigm there is a clear answer to (i), and a conjectured answer to (ii). For more details on fuzzballs see \cite{fuzzballs_i,fuzzballs_ii,fuzzballs_iii,fuzzballs_iv,fuzzballs_v}.

 For (i), we note that in the fuzzball paradigm the region $r\lesssim r_h + l_p$  is not described by a vacuum region but by a collection of horizon sized fuzzball states. Suppose an object of energy $E$ falls onto a black hole of mass $M$. Then the relevant radius $r_h$ is the horizon radius for a black hole of mass $M+E$. We expect that a  typical fuzzball state of this mass has a radius $\approx r_h+l_p$. As long as $r\gtrsim r_h+l_p$, the infalling object does not have a significant overlap with the fuzzballs of mass $M+E$. But as the object reaches $r\approx r_h+l_p$,  the overall gravity wavefunctional evolves so that the energy of the infalling object gets transformed into the nucleation of fuzzballs of mass $M+E$. 
 
 For (ii), we recall the conjecture of fuzzball complementarity. We have already noted that as the object reaches $r\approx r_h+l_p$ its energy  gets transformed into altering the wavefunctional of fuzzballs of mass $M+E$ from the vacuum state to an alternative state which we write schematically as $|F(t)\rangle$. The further evolution of the system must be understood as an evolution in this `superspace' -- the space of all fuzzball configurations. If the conjecture of fuzzball complementarity is true, then this evolution in superspace can be mapped, approximately, to infall in a traditional black hole geometry. The approximation gets better in the limit $E\gg T$, but fails at $E\lesssim T$. Thus the modes involved in Hawking radiation do not see any effective geometry like that of the traditional hole, and information of the fuzzball is carried out by such modes. It is important that only {\it infalling} modes with $E\gg T$ see the effective black hole interior; outgoing modes inside the hole have negative energy, and are not described by the approximation. 

In \cite{model} a model was given for the evolution in the space of fuzzball states, where an approximate description of infall could be deduced. In such a picture, thermalization happens when the perturbation caused by the infalling object has spread over all the accessible fuzzball states. This is because initially the wavefunctional spreads in a coherent way over superspace, but when it has spread over all the accessible space then it starts to decohere. This second stage is what we will call thermalization. 

To summarize, in our picture there are three stages:

\b

(i) In the first stage, in the gravity picture,  the infalling object just falls to the vicinity of the horizon. In the dual CFT this part of the evolution is no different from the CFT description of  infall in a geometry like (\ref{one}) which has no horizon. 

\b

(ii) In the gravity picture the object gets so close to the horizon that its energy starts getting converted to fuzzball states. There is then an evolution in the space of these fuzzball states.  If the conjecture of fuzzball complementarity is true, this evolution in the space of fuzzball states can be mapped,  approximately,  to a collective mode description that mimics infall from the horizon to the singularity. 

\b

(iii) In the gravity description the wavefunction over the space of fuzzballs has ended its coherent spread over fuzzball states and starts to become a  non-smooth function over superspace. This is the step which in the dual CFT should correspond to the process of thermalization. (If the conjecture of fuzzball complementarity is true, this step in the gravity picture corresponds to reaching the singularity.) 

\b

\subsection{The computation of this paper}

From the above discussion we see that thermalization starts when the wavefunction has completed its initial spread over accessible states. To see thermalization explicitly we will therefore have to take a system with a very small number of states. We take the gravity system to be that produced by a bound state of D1 and D5 branes.  Suppose we have $n_1$ D1 branes and $n_5$ D5 branes. Then the dual 1+1 dimensional CFT is characterized by a number
\be
N=n_1n_5
\ee
The CFT has $N$ copies of a $c=6$ CFT at its free `orbifold' point. We take $N=2$. This will allow us to explain the notion of what we should consider as thermalization in the CFT, but will obviously not map to a well defined black hole in the gravity picture. 

There is no thermalization at the orbifold point since the theory is free. We perturb away from the orbifold point by a $(1,1)$ deformation operator $D$. The perturbation theory in $D$ has been studied in a series of papers \cite{Avery:2010er,Burrington:2012yq,Burrington:2014yia,Carson:2016uwf,peet}. There is no clear evidence of thermalization at first order in the perturbation, but as we will see now, there is a 1-loop process at second order in $D$ which gives an effect that indicates thermalization. 

In the D1D5 CFT there are a set of ground states, which are characterized by the windings of the $N$ copies of the $c=6$ CFT. With $N=2$, there are two winding sectors: one where both copies are singly wound, and one where there is one doubly wound copy. A state with all copies singly wound (and all spins on these copies aligned) corresponds to  global AdS in the dual gravity theory. We start with such a state: i.e., we take two singly wound copies with their spins aligned. We now consider two kinds of excitations:

\b

(a) Consider the gravity picture. If we send in {\it one} particle from infinity into global AdS, then the particle  will travel along a geodesic and emerge to the other side of AdS without making a black hole; this happens because 
a single particle will travel along a geodesic, and by symmetry all points along its path are equivalent. In the dual CFT, we can consider a single particle excitation around the CFT state that described global AdS. We find that the the effect of the perturbation operator $D$ generates oscillations around the leading order state, but does not lead to a secular term where the perturbation continues to grow. We can therefore say that to this order the computation in the CFT agrees with the expectation from gravity. 

\b

(b) Consider the gravity picture, and now  send {\it two} particles into AdS from opposite directions.  These particles can collide and make a black hole if their total energy exceeds the threshold mass corresponding to the smallest AdS-Schwarzschild hole. The mass of such a hole grows with $N$,  so for $N=2$ the threshold  can be reached with quite a low excitation energy. In the dual CFT, we find that in this case the perturbation $D$ generates oscillations of the state as in case (a), but also gives secular terms that grow as $t^2$. We take this as evidence that if we apply the 1-loop perturbation several times, the state will drift in the space of all states, while the state in (a) did not have this behavior. 

\b

The above two computations therefore serve to illustrate the difference between a particle moving in AdS (where we have redshift) and a situation where a collision can lead to thermalization (where we have redshift as well as black hole formation). Since we are at a low value of $N$ and at only second order in the coupling $\lambda$, we do not have a good black hole in the dual gravity theory. Thus our computations are meant to be only suggestive of what kinds of effects should correspond to thermalization, rather than showing the details of black hole formation in the CFT. They do involve putting together several tools that have been developed in earlier papers: inserting two deformation operators $D$ in the CFT, and integrating over the positions of these operators.

\section{The orbifold CFT}\label{CFT}

Let us begin by recalling the orbifold CFT that we will be working with. Consider type IIB string theory, compactified as:
\bea
M_{9,1} &\to& M_{4,1}\times S^1\times T^4.
\eea
Now wrap $N_1$ D1 branes on $S^1$ and $N_5$ D5 branes on $S^1 \times T^4$.  We take $S^1$ to be large compared to $T^4$, so that the low energies are dominated by excitations only in the direction $S^1$.  This low-energy limit gives a $1+1$ dimensional CFT living on $S^1$.

At this point, variations in the moduli of string theory move us through the moduli space of the CFT on $S^1$.  It is conjectured that we can move to an 'orbifold point' where this CFT is a particularly simple sigma model \cite{orbifold2}. For many nice results using the D1D5 CFT at the `orbifold point' see \cite{orbifold2,cm,Larsen:1999uk,Arutyunov:1997gt,dmw,sv,lm1,lm2,taylor}. 
  We will begin in the Euclidean theory at this orbifold point.  The base space is a cylinder spanned by the coordinates $\t,\s$:
\bea
0\leq\s<2\pi,\qquad -\infty<\t<\infty.
\label{range}
\eea
These coordinates can be collected into a single complex coordinate
\bea
w = \t + i\s
\label{cylinder coordinate}
\eea
The target space of this CFT is the symmetrized product of $N_1 N_5$ copies of $T^4$:
\bea
(T^4)^{N_1 N_5}/S_{N_1 N_5}.
\eea
Each copy gives 4 bosonic excitations and 4 fermionic excitations.  With an index $i$ ranging from 1 to 4, we label the bosonic excitations $X^i$, the left-moving fermionic excitations $\psi^i$, and the right-moving fermionic excitations $\bar{\psi}^i$.  The total central charge is then $6 N_1 N_5$. 

\subsection{The deformation operator}\label{sebsec_deformation}
The orbifold CFT describes the system at its `free' point in moduli space. To move towards the supergravity description, we deform the orbifold CFT by adding a deformation operator $D$.


To understand the structure of $D$ we recall that the orbifold CFT contains `twist' operators. Twist operators can link any number $k$ out of the $N$ copies of the CFT together to give a $c=6$ CFT living on a circle of length $2\pi k$ rather than $2\pi$.    We will call such a set of linked copies a `component string' with winding number $k$.

The deformation operator contains a twist of order $2$. The twist itself carries left and right charges $j=\pm \h, \bar j=\pm \h$ \cite{lm2}. Suppose we start with both these charges positive; this gives the twist $\sigma_2^{++}$. Then the deformation operators in this twist sector have the form
\be\label{exactlymarginal}
D=P^{\dot A\dot B}\hat O_{\dot A\dot B}= P^{\dot A \dot B}G^-_{\dot A, -\h}\bar G^-_{\dot B, -\h} \sigma^{++}_2\ .
\ee
Here $P^{\dot A \dot B}$ is a polarization. We will later choose
\be
P^{\dot A \dot B}=\epsilon^{\dot A \dot B}
\ee
where $\epsilon^{+-}=-1$. This choice gives a deformation carrying no charges.
The details of the deformation operator are given in \cite{Avery:2010er,Carson:2016uwf}. 


\subsection{The Amplitude at Second Order}\label{sebsec_deformation}

In this section we discuss the amplitudes we wish to compute using the deformation operator (\ref{exactlymarginal}).
We begin by writing the action of a perturbed CFT:
\be
S_0\rightarrow S_{\mathrm{pert}}=S_0+\lambda \int d^2 w D(w, \bar w)\ ,
\ee
with $D$ given in (\ref{exactlymarginal}).
Since we are interested in second order effects, the amplitude we wish to compute is of the form
\bea
\mathcal{A}^{i\to f}_{\text{int}} 
&=&{1\over2}\lambda^2\int d^2w_2\int d^2w_1\langle \Phi_f |D(w_2,\bar{w}_2)D(w_1,\bar{w}_1)|\Psi_i\rangle
\label{amplitude}
\eea
where $|\Psi_i\rangle,\langle\Phi_j|$ represent various initial and final states labeled by $i,j$ respectively, which we choose in the next section.
The states $|\Psi_i\rangle,\langle\Phi_j|$ include both holomorphic and antiholomorphic components.

\section{The states and amplitudes}\label{states amplitudes}

In this section we list the initial and final states that we will use in the amplitude (\ref{amplitude}). 

\subsection{The  initial states in the CFT}

We have a set of D1 and D5 branes making a bound state. Let the ground state of these branes be the NS sector vacuum $|0\rangle$. The dual spacetime is $AdS_3\times S^3\times T^4$. 

Let the $T^4$ be described by the indices  $I,J=1, \dots 4$. Consider gravitons in the gravity theory with indices along the $T^4$. Such a  graviton is described by a transverse traceless tensor $h_{IJ}$. If we do a dimensional reduction on the $T^4$, then such gravitons give minimally coupled scalars in the resulting $5+1$ dimensional gravity theory. 

In the orbifold CFT, the operator dual to this graviton is  
\be
O^h_{IJ}\sim \h \sum_{k=1}^N\left ( \partial_z X^{(k)}_I\partial_{\bar z} X^{(k)}_J+ \partial_z X^{(k)}_J\partial_{\bar z} X^{(k)}_I\right )
\ee
Here $(k)$ is an index labelling the $N$ copies of the $c=6$ CFT. Note that both the left and the right moving excitations appear on the {\it same} copy, and then we have a uniform sum over all the copies. Since the graviton is symmetric in its indices $I,J$, the operator is symmetric as well. 

If we antisymmetrize rather than symmetrize in the indices $I,J$, then the operator corresponds to  a quantum of the RR field $B^{RR}_{IJ}$, again with indices along the $T^4$ directions. Finally, the trace in the $I,J$ indices corresponds to absorbing the dilaton $\phi$. 

Thus any operator of the form
\be
O_{IJ}\sim  \sum_{k=1}^N \partial_z X^{(k)}_I\partial_{\bar z} X^{(k)}_J
\label{lone}
\ee
corresponds to some linear combination of $h_{IJ}, B_{IJ}^{RR}, \phi$. Thus we can use these operators $O_{IJ}$ as a convenient basis to study the dynamics of the simplest supergravity fields in the $AdS$ dual.

The indices $I, J, \dots$ describe a vector representation of $SO(4)\approx SU(2)\times SU(2)$. Decomposing the vector into two spinors we get the indices $I\rightarrow A\dot A$ where $A=\{ +, -\}$ and $\dot A=\{ +, -\}$. 

Thus the bosonic oscillators  will be described by operators $\alpha_{A\dot A, -n}$. We add one other index to these operators: a superscript that could be $i$ (for initial state) or $f$ (for final state). In general the winding sector $\{ k_1, k_2, \dots \} $ of the  CFT changes when we apply twist operators. Therefore the bosonic and fermionic modes in the initial and final states  are defined on different twist sectors, and so are not given by the same oscillators $\alpha, d$. In the cases that we will consider, the initial and final twist sectors happen to be the same. It is nevertheless useful to keep the superscripts $i,f$ since this index will help keep track of the position of the operator on the covering space $t$ when we undo the twists.

 We wish to start with  empty $AdS_3$; the corresponding state in the CFT is the NS vacuum $|0\rangle$. Since we are taking $N=2$, this vacuum is
\be
|0\rangle = |0\rangle^{(1)}|0\rangle^{(2)}|\bar{0}\rangle^{(1)}|\bar{0}\rangle^{(2)}
\ee
where the superscripts $(1), (2) $ describe the two copies and the unbarred and barred states are for the left and right moving sectors respectively. 

We consider two situations:

\b

(a) We have a single graviton sent into the $AdS$. This corresponds to an initial mode that is composed of one left bosonic oscillator and one right bosonic oscillator. We consider the state
\bea
|\Psi_1\rangle &=& {1\over \sqrt2}{1\over n}\a^{(1)i}_{--,-n}|0\rangle^{(1)}|0\rangle^{(2)}\bar{\a}^{(1)i}_{++,-n}|\bar{0}\rangle^{(1)}|\bar{0}\rangle^{(2)}\cr
&& +  {1\over \sqrt2}{1\over n}\a^{(2)i}_{--,-n}|0\rangle^{(1)}|0\rangle^{(2)}\bar{\a}^{(2)i}_{++,-n}|\bar{0}\rangle^{(1)}|\bar{0}\rangle^{(2)}\nn
\label{single excitation}
\eea
Note that we place  the left and right oscillators on the same copy of the CFT   and then sum over the copies with equal weight; this is required by the form of the vertex operator (\ref{lone}). The state (\ref{single excitation}) has the quantum numbers
\bea
h = n,~~j=0;\qquad \bar{h}=n,~~\bar{j}=0
\eea
By superposing such states for different $n$, we can describe a graviton that is sent in from the boundary of $AdS$ into the interior. 

\b

(b) We send two gravitons into the $AdS$. The initial state is then
\bea
|\Psi_2\rangle &=& {1\over2}{1\over n_1n_2}\a^{(1)i}_{++,-n_1}\a^{(1)i}_{--,-n_2}|0\rangle^{(1)}|0\rangle^{(2)}\bar{\a}^{(1)i}_{--,-n_1}\bar{\a}^{(1)i}_{++,-n_2}|\bar{0}\rangle^{(1)}|\bar{0}\rangle^{(2)} \cr
&&\!\!\!\!+ {1\over2}{1\over n_1n_2}\a^{(1)i}_{++,-n_1}\a^{(2)i}_{--,-n_2}|0\rangle^{(1)}|0\rangle^{(2)}  \bar{\a}^{(1)i}_{--,-n_1}\bar{\a}^{(2)i}_{++,-n_2}|\bar{0}\rangle^{(1)}|\bar{0}\rangle^{(2)}\cr
&&\!\!\!\!+ {1\over2}{1\over n_1n_2}\a^{(2)i}_{++,-n_1}\a^{(1)i}_{--,-n_2}|0\rangle^{(1)}|0\rangle^{(2)}  \bar{\a}^{(2)i}_{--,-n_1}\bar{\a}^{(1)i}_{++,-n_2}|\bar{0}\rangle^{(1)}|\bar{0}\rangle^{(2)}\cr
&&\!\!\!\!+ {1\over2}{1\over n_1n_2}\a^{(2)i}_{++,-n_1}\a^{(2)i}_{--,-n_2}|0\rangle^{(1)}|0\rangle^{(2)}  \bar{\a}^{(2)i}_{--,-n_1}\bar{\a}^{(2)i}_{++,-n_2}|\bar{0}\rangle^{(1)}|\bar{0}\rangle^{(2)}
\label{initial states}
\eea
The state (\ref{initial states}) has the quantum numbers
\bea
h=n_1 + n_2,~~j=0;\qquad \bar{h} = n_1 + n_2,~~\bar{j}=0
\eea
By superposing such states for different $n_1, n_2$, we can describe a pair of gravitons that are sent in from opposite sides of the boundary of $AdS$ into the interior. 

\subsection{The final states in the CFT}

If we are at the orbifold point then the initial states defined above stay unchanged with time. But if we consider the interaction given by the deformation operator $D$, then the states evolve. We are starting with two untwisted (i.e., singly wound) copies of the CFT. The twist in $D$ can join the two unwound copies into a doubly wound copy, and the twist in a second $D$  can split these into two singly wound copies again. The state of the bosonic and fermionic oscillators will not, however, return to its initial one.  In particular, suppose the initial state had a high energy; this is given by large $n\gg 1$ in (\ref{single excitation})  and large $n_1, n_2\gg 1$ in (\ref{initial states}). Then there is a high probability for the initial energy to be split among several excitations.

The CFT amplitudes that we will encounter factorize into a holomorphic (left) factor and a antiholomorphic (right) factor. There is no particular reason for the final state to be the same in these two sectors. But since our goal is to explore some examples of the states generated by interactions, we do take the left and right sides to split into multiple excitations in the {\it same} way. For the two initial states mentioned above, we consider the following final states:

\b

(a') Consider the initial state (\ref{single excitation}), and focus on the left sector.  There is one bosonic  oscillator $\alpha$. After two actions of the deformation operator $D$, we can end up with one oscillator excitation, or three excitations or five excitations, and so on; these possibilities are dictated by the action of the twists and the fact that there is a supercharge $G$ acting at the twists. Since we are looking at how the initial energy is split between modes, we ignore the case when we have just one oscillator in the final state, and focus on the case where we have three excitations in the final state. There are two possibilities: we can end up with three bosons  or one boson and two fermions. As mentioned above, for simplicity we will take the right movers to split in exactly the same way as the left movers. Then we have the following two choices for our final states:

\b

(i) 3 Bosons:
\bea
\langle \Phi_1|&=&{1\over\sqrt{2}}{1\over pqr}{}^{(1)}\langle \bar{0}|{}^{(2)}\langle \bar{0}|\bar{\a}^{(1)f}_{--,p}\bar{\a}^{(1)f}_{++,q}\bar{\a}^{(1)f}_{--,r}{}^{(1)}\langle 0|{}^{(2)}\langle 0|\a^{(1)f}_{++,p}\a^{(1)f}_{--,q}\a^{(1)f}_{++,r}\cr
&&+{1\over\sqrt{2}}{1\over pqr}{}^{(1)}\langle \bar{0}|{}^{(2)}\langle \bar{0}|\bar{\a}^{(2)f}_{--,p}\bar{\a}^{(2)f}_{++,q}\bar{\a}^{(2)f}_{--,r}{}^{(1)}\langle 0|{}^{(2)}\langle 0|\a^{(2)f}_{++,p}\a^{(2)f}_{--,q}\a^{(2)f}_{++,r}
\eea

\b

(ii) 1 Boson 2 Fermions

\bea
\langle \Phi_2|&=&{1\over\sqrt{2}}{1\over p}{}^{(1)}\langle \bar{0}|{}^{(2)}\langle \bar{0}|\bar{\a}^{(1)f}_{--,p}\bar{d}^{(1)f,-+}_q\bar{d}^{(1)f,+-}_r{}^{(1)}\langle 0|{}^{(2)}\langle 0|\a^{(1)f}_{++,p}d^{(1)f,+-}_qd^{(1)f,-+}_r\cr
&&+{1\over\sqrt{2}}{1\over p}{}^{(1)}\langle \bar{0}|{}^{(2)}\langle \bar{0}|\bar{\a}^{(2)f}_{--,p}\bar{d}^{(2)f,-+}_q\bar{d}^{(2)f,+-}_r{}^{(1)}\langle 0|{}^{(2)}\langle 0|\a^{(2)f}_{++,p}d^{(1)f,+-}_qd^{(1)f,-+}_r
\eea

\b

(b')  Consider the initial state (\ref{initial states}), and again focus on the left sector. This time the number of excitations in the final state  can be  two, or four, or six and so on. We focus on the case where the energy splits among four excitations. There are two cases: we can get 4 bosons, or 2 bosons and 2 fermions.  Again, for simplicity, we take the right sector to have the same structure as the left sector. Thus we consider the following two final states:

\b

(i) 4 Bosons:
\bea
\langle \Phi_3|&=&{1\over\sqrt{2}}{1\over pqrs}{}^{(1)}\langle \bar{0}|{}^{(2)}\langle \bar{0}|\bar{\a}^{(1)f}_{--,p}\bar{\a}^{(1)f}_{++,q}\bar{\a}^{(1)f}_{--,r}\bar{\a}^{(1)f}_{++,s}\cr
&&\qquad\qquad\qquad{}^{(1)}\langle 0|{}^{(2)}\langle 0|\a^{(1)f}_{++,p}\a^{(1)f}_{--,q}\a^{(1)f}_{++,r}\a^{(1)f}_{--,s}\cr
&&+{1\over\sqrt{2}}{1\over pqrs}{}^{(1)}\langle \bar{0}|{}^{(2)}\langle \bar{0}|\bar{\a}^{(2)f}_{--,p}\bar{\a}^{(2)f}_{++,q}\bar{\a}^{(2)f}_{--,r}\bar{\a}^{(2)f}_{++,s}\cr
&&\qquad\qquad\qquad{}^{(1)}\langle 0|{}^{(2)}\langle 0|\a^{(2)f}_{++,p}\a^{(2)f}_{--,q}\a^{(2)f}_{++,r}\a^{(2)f}_{--,s}
\eea

\b

(ii) 2 Bosons, 2 Fermions:
\bea
\langle \Phi_4|&=&{1\over\sqrt{2}}{1\over pq}{}^{(1)}\langle \bar{0}|{}^{(2)}\langle \bar{0}|\bar{\a}^{(1)f}_{--,p}\bar{\a}^{(1)f}_{++,q}\bar{d}^{(1)f,-+}_r\bar{d}^{(1)f,+-}_s\cr
&&\qquad\qquad\qquad{}^{(1)}\langle 0|{}^{(2)}\langle 0|\a^{(1)f}_{++,p}\a^{(1)f}_{--,q}d^{(1)f,+-}_rd^{(1)f,-+}_s\cr
&&+{1\over\sqrt{2}}{1\over pq}{}^{(1)}\langle \bar{0}|{}^{(2)}\langle \bar{0}|\bar{\a}^{(2)f}_{--,p}\bar{\a}^{(2)f}_{++,q}\bar{d}^{(2)f,-+}_r\bar{d}^{(2)f,+-}_s\cr
&&\qquad\qquad\qquad{}^{(1)}\langle 0|{}^{(2)}\langle 0|\a^{(2)f}_{++,p}\a^{(2)f}_{--,q}d^{(1)f,+-}_rd^{(1)f,-+}_s\nn
\label{final states}
\eea

\subsection{The quantities to be computed}

We need to compute the amplitudes where we start in one of our initial states, apply two deformation operators, and end up in one of our chosen allowed final states. First consider the left sector, and let the deformation operators be at positions $w_1, w_2$. We need to compute the following four amplitudes:
\bea
&&\!\!\!\!\!\!\!\!\!\!\!\!\!\!\!\!\!\!\!\!\mathcal{A}^{\a\to \a\a\a}(w_2,w_1,\bar{w}_2,\bar{w}_1)\cr
&&=\langle \Phi_1|D(w_2,\bar w_2)D(w_1,\bar w_1)|\Psi_1\rangle\cr
&&=e^{\dot{C}\dot{D}}\e^{\dot{A}\dot{B}}\langle \Phi_1|\big(G^+_{\dot{C},-{1\over2}}\s^-(w_2)G^-_{\dot{A},-{1\over2}}\s^+(w_1)\big)\big(\bar{G}^+_{\dot{D},-{1\over2}}\bar{\s}^-(\bar{w}_2)\bar{G}^-_{\dot{B},-{1\over2}}\bar{\s}^+(\bar{w}_1)\big)|\Psi_1\rangle
\cr
\cr
&&\!\!\!\!\!\!\!\!\!\!\!\!\!\!\!\!\!\!\!\!\mathcal{A}^{\a\to \a dd}(w_2,w_1,\bar{w}_2,\bar{w}_1)\cr
&&=\langle \Phi_2|D(w_2,\bar w_2)D(w_1,\bar w_1)|\Psi_1\rangle\cr
&&=\e^{\dot{C}\dot{D}}\e^{\dot{A}\dot{B}}\langle \Phi_2|\big(G^+_{\dot{C},-{1\over2}}\s^-(w_2)G^-_{\dot{A},-{1\over2}}\s^+(w_1)\big)\big(\bar{G}^+_{\dot{D},-{1\over2}}\bar{\s}^-(\bar{w}_2)\bar{G}^-_{\dot{B},-{1\over2}}\bar{\s}^+(\bar{w}_1)\big)|\Psi_1\rangle
\cr
\cr
&&\!\!\!\!\!\!\!\!\!\!\!\!\!\!\!\!\!\!\!\!\mathcal{A}^{\a\a\to \a\a \a\a}(w_2,w_1,\bar{w}_2,\bar{w}_1)\cr
&&=\langle \Phi_3|D(w_2,\bar w_2)D(w_1,\bar w_1)|\Psi_2\rangle\cr
&&=\e^{\dot{C}\dot{D}}\e^{\dot{A}\dot{B}}\langle \Phi_3|\big(G^+_{\dot{C},-{1\over2}}\s^-(w_2)G^-_{\dot{A},-{1\over2}}\s^+(w_1)\big)\big(\bar{G}^+_{\dot{D},-{1\over2}}\bar{\s}^-(\bar{w}_2)\bar{G}^-_{\dot{B},-{1\over2}}\bar{\s}^+(\bar{w}_1)\big)|\Psi_2\rangle
\cr
\cr
&&\!\!\!\!\!\!\!\!\!\!\!\!\!\!\!\!\!\!\!\!\mathcal{A}^{\a\a\to \a\a dd}(w_2,w_1,\bar{w}_2,\bar{w}_1)\cr
&&=\langle \Phi_4|D(w_2,\bar w_2)D(w_1,\bar w_1)|\Psi_2\rangle\cr
&&=\e^{\dot{C}\dot{D}}\e^{\dot{A}\dot{B}}\langle \Phi_4|\big(G^+_{\dot{C},-{1\over2}}\s^-(w_2)G^-_{\dot{A},-{1\over2}}\s^+(w_1)\big)\big(\bar{G}^+_{\dot{D},-{1\over2}}\bar{\s}^-(\bar{w}_2)\bar{G}^-_{\dot{B},-{1\over2}}\bar{\s}^+(\bar{w}_1)\big)|\Psi_2\rangle
\cr\nn
\label{all amplitudes}
\eea

We must integrate over the positions of the deformation operators. We assume that the initial state is at $t_i=-{t\over2}$ and the final state is at $t_f={t\over2}$. We integrate the positions of the deformation operators in the interval in between. This step is shown in Section \ref{integration}. 
Next we summarize the steps we took to compute these splitting amplitudes.

\subsection{Procedure for Computing Amplitudes}

In this subsection, we summarize the various steps for computing the final splitting amplitudes that are presented in this paper. The details of the computation are not give explicitly here;  they are lengthy but straightforward if we use the techniques in the papers mentioned below. In addition, a full and explicit derivation of the results  can be found  in \cite{dissertation}. 

The steps are as follows:

\b

\begin{enumerate}
\item The amplitudes in \ref{all amplitudes} are defined on a base space which has the shape of a cylinder. The coordinate on the cylinder is $w$ (\ref{cylinder coordinate}), with range given by  (\ref{range}). For details about the D1D5 CFT and the deformation $D$ see \cite{Avery:2010er,Carson:2016uwf,dissertation}. 

\item The operator insertions on the cylinder involve twist operators, which makes the fields multi-valued functions of the cylinder coordinate $w$. We wish to pass to  a  covering space where all fields will be single valued. To do this, we first map the cylinder   to the complex plane through the map 
\bea
z=e^w
\eea 
We then pass to from the $z$ plane   to the covering $t$ plane through the map 
\bea
z={(t+a)(t+b)\over t}.
\eea
For details regarding this map see \cite{Carson:2016uwf}. The fields defined in the $t$ plane are now single valued. 

\item  When mapping to the $t$ plane our amplitudes in \ref{all amplitudes} take the schematic form
\bea
\mathcal{A}^{i\to f}\sim C \mathcal{A}^{i\to f}_t.
\eea
Here the function $C$ contains the Jacobians arising from the changes of coordinates from $w$ to $t$.  These Jacobians are given in detail \cite{dissertation,lifting}. In the $t$ plane we now have single valued bosonic and fermionic fields, and various operator insertions of these fields. The factor $\mathcal{A}^{i\to f}_t$ arises from the various $t$ plane Wick contractions between pairs of these bosonic and fermionic operators. The techniques used to compute these contractions are given in \cite{Carson:2016uwf}. These techniques are combined to compute $\mathcal{A}^{i\to f}_t$ which is given in detail in \cite{dissertation}.

\item The above procedure gives a holomorphic part of the amplitude and an antiholomorphic part. These parts are multiplied together, and then the positions of the deformation operators are integrated over. 

\end{enumerate}

\b

We now give the results of these steps, finally evaluating the results for specific choices of mode numbers. The result for the general case can be expressed as a finite sum of terms, but it is not easy to see the physical nature of the interaction from such expressions.

\section{Computing our amplitudes}\label{computing amplitudes}

Consider the amplitudes  (\ref{all amplitudes}), and let them be multiplied by their right moving complex conjugate factors. After we integrate the positions of the deformation operators over the $\sigma$ circle, we will find that the action of the deformation cannot change the momentum of the state. We will choose our final state to have the same energy as the initial state. In this case the left and right moving levels are unchanged by the action of the two deformation operators. The amplitudes (\ref{all amplitudes}) are then a function of 
\be
\Delta w=w_2 - w_1; ~~~~~\Delta \bar w = \bar w_2-\bar w_1
\ee
The full amplitudes  will thus take the schematic form
\bea
\mathcal{A}^{i\to f}(w_1,w_2,\bar w_1,\bar w_2)=\sum_{m=m_{min}(n_{i};n_{f})}^{m_{max}(n_{i};n_{f})}\sum_{m'=m'_{min}(n_{i};n_{f})}^{m'_{max}(n_{i};n_{f})}B^{i\to f}_{m,m'}(n_i;n_f) e^{{m\Delta \bar w\over2}+{m'\Delta w\over2}}
\label{ltwo}
\eea
Here the symbol  $i\to f$ denotes the specific  process in consideration; i.e., one of the four cases
\be
\a\to \a\a\a, ~~~
\a\to \a dd, ~~~
 \a\a\to\a\a\a\a, ~~~
 \a\a\to \a\a dd
\ee
 The $B^{i\to f}_{m,m'}(n_i;n_f)$ are numerical coefficients that we must find. The numbers $m, m'$ range over a finite set of integers, with lower and upper bounds as specified in the summations. The symbols $n_i, n_f$ denote the set of  mode numbers for the initial and final states respectively.

\subsection{The amplitude for $\a\to \a\a\a$}

Consider the  amplitude for one boson going to three bosons on the left, and the same split occurring on the right. We take the following initial and final mode numbers
\bea
n_i&=&\lbrace n=3\rbrace\cr
n_f &=& \lbrace p=1,q=1,r=1\rbrace
\eea
Then we find \cite{dissertation}
\bea
&&\!\!\!\!\!\!\!\!\!\!\!\!\mathcal{A}^{\a\to \a\a\a}(w_1,w_2,\bar w_1,\bar w_2)\cr 
&&\!\!\!\!\!\!=e^{\dot{C}\dot{D}}\e^{\dot{A}\dot{B}}\langle \Phi_3|\big(G^+_{\dot{C},-{1\over2}}\s^-(w_2)G^-_{\dot{A},-{1\over2}}\s^+(w_1)\big)\big(\bar{G}^+_{\dot{D},-{1\over2}}\bar{\s}^-(\bar{w}_2)\bar{G}^-_{\dot{B},-{1\over2}}\bar{\s}^+(\bar{w}_1)\big)|\Psi_2\rangle
\cr
&&\!\!\!\!\!\!= \frac{75 e^{-\frac{5 \Delta w}{2}-\frac{5 \Delta \bar w}{2}}}{131072}-\frac{15 e^{-\frac{3 \Delta w}{2}-\frac{5 \Delta \bar w}{2}}}{65536}-\frac{15e^{-\frac{\Delta w}{2}-\frac{5 \Delta \bar w}{2}}}{32768}-\frac{45 e^{\frac{\Delta w}{2}-\frac{5 \Delta \bar w}{2}}}{65536}+\frac{105 e^{\frac{3 \Delta w}{2}-\frac{5\Delta \bar w}{2}}}{131072}
   \cr
   &&-\frac{15 e^{-\frac{5 \Delta w}{2}-\frac{3 \Delta \bar w}{2}}}{65536}+\frac{159 e^{-\frac{3 \Delta w}{2}-\frac{3 \Delta \bar w}{2}}}{131072}-\frac{51e^{-\frac{\Delta w}{2}-\frac{3 \Delta \bar w}{2}}}{65536}-\frac{3 e^{\frac{\Delta w}{2}-\frac{3 \Delta \bar w}{2}}}{8192}-\frac{21 e^{\frac{3 \Delta w}{2}-\frac{3\Delta \bar w}{2}}}{32768}
   \cr
   &&+\frac{9e^{\frac{\Delta w}{2}-\frac{\Delta \bar w}{2}}}{8192}-\frac{3 e^{\frac{3 \Delta w}{2}-\frac{\Delta \bar w}{2}}}{8192}-\frac{45 e^{\frac{5
   \Delta w}{2}-\frac{\Delta \bar w}{2}}}{65536}-\frac{45 e^{-\frac{5 \Delta w}{2}+\frac{\Delta \bar w}{2}}}{65536}-\frac{3 e^{-\frac{3 \Delta w}{2}+\frac{\Delta \bar w}{2}}}{8192}+\frac{9
   e^{-\frac{\Delta w}{2}+\frac{\Delta \bar w}{2}}}{8192}\cr
   &&+\frac{39 e^{\frac{\Delta w}{2}+\frac{\Delta \bar w}{2}}}{32768}+\frac{105 e^{\frac{5 \Delta w}{2}-\frac{3 \Delta \bar w}{2}}}{131072}-\frac{15 e^{-\frac{5 \Delta w}{2}-\frac{\Delta \bar w}{2}}}{32768} -\frac{51e^{-\frac{3 \Delta w}{2}-\frac{\Delta \bar w}{2}}}{65536}+\frac{39 e^{-\frac{\Delta w}{2}-\frac{\Delta \bar w}{2}}}{32768}
   \cr
   &&-\frac{51 e^{\frac{3\Delta w}{2}+\frac{\Delta \bar w}{2}}}{65536}-\frac{15 e^{\frac{5 \Delta w}{2}+\frac{\Delta \bar w}{2}}}{32768}+\frac{105 e^{-\frac{5
   \Delta w}{2}+\frac{3 \Delta \bar w}{2}}}{131072}-\frac{21 e^{-\frac{3 \Delta w}{2}+\frac{3 \Delta \bar w}{2}}}{32768}-\frac{3 e^{-\frac{\Delta w}{2}+\frac{3 \Delta \bar w}{2}}}{8192}
   \cr
   &&-\frac{51e^{\frac{\Delta w}{2}+\frac{3 \Delta \bar w}{2}}}{65536} + \frac{159 e^{\frac{3 \Delta w}{2}+\frac{3 \Delta \bar w}{2}}}{131072}-\frac{15 e^{\frac{5 \Delta w}{2}+\frac{3\Delta \bar w}{2}}}{65536}+\frac{105 e^{-\frac{3 \Delta w}{2}+\frac{5 \Delta \bar w}{2}}}{131072}-\frac{45 e^{-\frac{\Delta w}{2}+\frac{5 \Delta \bar w}{2}}}{65536}\cr
   &&-\frac{15e^{\frac{\Delta w}{2}+\frac{5 \Delta \bar w}{2}}}{32768}-\frac{15 e^{\frac{3 \Delta w}{2}+\frac{5 \Delta \bar w}{2}}}{65536}+\frac{75 e^{\frac{5 \Delta w}{2}+\frac{5
   \Delta \bar w}{2}}}{131072}
   \label{a to aaa 3 to 111}
\eea
Writing (\ref{a to aaa 3 to 111}) in the form (\ref{ltwo})  we have
\bea
\mathcal{A}^{\a\to \a\a\a}(w_1,w_2,\bar w_1,\bar w_2) = \sum_{m=-5}^{5} \sum_{m'=-5}^{5}B^{\a\to \a\a\a}_{m,m'}(n=3; p=1,q=1,r=1)e^{{m\Delta w\over2}+{m'\Delta \bar w\over2}}\nn
\eea
This defines the coefficients $B^{\a\to \a\a\a}_{m,m'}(n=3; p=1,q=1,r=1)$.


\subsection{The amplitude for $\a\to \a dd$}

Consider the  amplitude for one boson going to one boson and two fermions on the left, and the same split occurring on the right. We take the following initial and final mode numbers
\bea
n_i&=&\lbrace n=2\rbrace\cr
n_f &=& \lbrace p=1,q={1\over2},r={1\over2}\rbrace
\eea
We find \cite{dissertation}
\bea
&&\!\!\!\!\!\!\!\!\!\!\!\!\mathcal{A}^{\a\to \a dd}(w_1,w_2,\bar w_1,\bar w_2)\cr
&&=\e^{\dot{C}\dot{D}}\e^{\dot{A}\dot{B}}\langle \Phi_2|\big(G^+_{\dot{C},-{1\over2}}\s^-(w_2)G^-_{\dot{A},-{1\over2}}\s^+(w_1)\big)\big(\bar{G}^+_{\dot{D},-{1\over2}}\bar{\s}^-(\bar{w}_2)\bar{G}^-_{\dot{B},-{1\over2}}\bar{\s}^+(\bar{w}_1)\big)|\Psi_2\rangle
\cr
&& = \frac{9 e^{-\frac{3\Delta w}{2}-\frac{3 \Delta  \bar{w}}{2}}}{4096}-\frac{3e^{-\frac{\Delta w}{2}-\frac{3 \Delta  \bar{w}}{2}}}{2048}-\frac{3 e^{\frac{\Delta w}{2}-\frac{3 \Delta  \bar{w}}{2}}}{4096}-\frac{3 e^{-\frac{3 \Delta w}{2}-\frac{\Delta  \bar{w}}{2}}}{2048}+\frac{5 e^{-\frac{\Delta w}{2}-\frac{\Delta  \bar{w}}{2}}}{4096}+\frac{e^{\frac{\Delta w}{2}-\frac{\Delta  \bar{w}}{2}}}{1024}
\cr
&&\quad -\frac{3e^{\frac{3\Delta w}{2}-\frac{\Delta  \bar{w}}{2}}}{4096}-\frac{3 e^{-\frac{3\Delta w}{2}+\frac{\Delta \bar{w}}{2}}}{4096}+\frac{e^{-\frac{\Delta w}{2}+\frac{\Delta \bar{w}}{2}}}{1024}+\frac{5 e^{\frac{\Delta w}{2}+\frac{\Delta \bar{w}}{2}}}{4096}-\frac{3 e^{\frac{3\Delta w}{2}+\frac{\Delta  \bar{w}}{2}}}{2048} -\frac{3 e^{-\frac{\Delta w}{2}+\frac{3 \Delta  \bar{w}}{2}}}{4096}\cr
&&\quad-\frac{3 e^{\frac{\Delta w}{2}+\frac{3 \Delta  \bar{w}}{2}}}{2048}+\frac{9e^{\frac{3\Delta w}{2} + \frac{3 \Delta  \bar{w}}{2}}}{4096} 
\label{a to add 2 to 1halfhalf}
\eea
We write this as
\bea
\mathcal{A}^{\a\to \a dd}(w_1,w_2,\bar w_1,\bar w_2) = \sum_{m=-3}^{3} \sum_{m'=-3}^{3}B^{\a\to \a dd}_{m,m'}(n=2; p=1,q={1\over2},r={1\over2})e^{{m  \Delta  w\over2}+{m'\Delta \bar w\over2}}\nn
\label{a to add condensed}
\eea
which defines the coefficients $B^{\a\to \a dd}_{m,m'}(n=2; p=1,q={1\over2},r={1\over2})$.

\subsection{The amplitude for the process $\a\a\to \a\a \a\a$}

Consider the  amplitude for two bosons going to four bosons on the left, and the same split occurring on the right. We take the following initial and final mode numbers
\bea
n_i&=&\lbrace n_1=2,n_2=2\rbrace\cr
n_f &=& \lbrace p=1,q=1,r=1,s=1\rbrace
\eea
We find \cite{dissertation}
\bea
&&\!\!\!\!\!\!\!\!\!\!\!\!\mathcal{A}^{\a\a\to \a\a \a\a}(w_1,w_2,\bar w_1,\bar w_2)\cr
&&\!\!\!\!\!\!=\e^{\dot{C}\dot{D}}\e^{\dot{A}\dot{B}}\langle \Phi_4|\big(G^+_{\dot{C},-{1\over2}}\s^-(w_2)G^-_{\dot{A},-{1\over2}}\s^+(w_1)\big)\big(\bar{G}^+_{\dot{D},-{1\over2}}\bar{\s}^-(\bar{w}_2)\bar{G}^-_{\dot{B},-{1\over2}}\bar{\s}^+(\bar{w}_1)\big)|\Psi_2\rangle
\cr
&&\!\!\!\!\!\!=\frac{59049}{8388608 \sqrt{2}}+\frac{37179 e^{-3 \Delta w}}{33554432\sqrt{2}}-\frac{29889 e^{-2 \Delta w}}{16777216 \sqrt{2}}-\frac{95499 e^{-\Delta w}}{33554432 \sqrt{2}}-\frac{95499 e^{\Delta w}}{33554432 \sqrt{2}}\cr
&&-\frac{29889 e^{2\Delta w}}{16777216 \sqrt{2}}+\frac{37179 e^{3 \Delta w}}{33554432\sqrt{2}}+\frac{37179 e^{-3 \Delta  \bar{w}}}{33554432 \sqrt{2}}-\frac{29889 e^{-2 \Delta \bar{w}}}{16777216 \sqrt{2}}-\frac{95499 e^{-\Delta  \bar{w}}}{33554432 \sqrt{2}}\cr
&&-\frac{95499e^{\Delta  \bar{w}}}{33554432 \sqrt{2}}-\frac{29889 e^{2 \Delta  \bar{w}}}{16777216\sqrt{2}}+\frac{37179 e^{3 \Delta  \bar{w}}}{33554432 \sqrt{2}}+\frac{23409 e^{-3 \Delta w-3 \Delta  \bar{w}}}{134217728 \sqrt{2}}\cr
&&-\frac{18819 e^{-2 \Delta w-3 \Delta\bar{w}}}{67108864 \sqrt{2}}-\frac{60129 e^{-\Delta w-3 \Delta  \bar{w}}}{134217728\sqrt{2}}-\frac{60129 e^{\Delta w-3 \Delta  \bar{w}}}{134217728 \sqrt{2}}-\frac{18819 e^{2\Delta w-3 \Delta  \bar{w}}}{67108864 \sqrt{2}}\cr
&&+\frac{23409 e^{3 \Delta w-3 \Delta\bar{w}}}{134217728 \sqrt{2}}-\frac{18819 e^{-3 \Delta w-2 \Delta  \bar{w}}}{67108864\sqrt{2}}+\frac{15129 e^{-2 \Delta w-2 \Delta  \bar{w}}}{33554432 \sqrt{2}}+\frac{48339e^{-\Delta w-2 \Delta  \bar{w}}}{67108864 \sqrt{2}}\cr
&&+\frac{48339 e^{\Delta w-2\Delta  \bar{w}}}{67108864 \sqrt{2}}+\frac{15129 e^{2 \Delta w-2 \Delta  \bar{w}}}{33554432\sqrt{2}}-\frac{18819 e^{3 \Delta w-2 \Delta  \bar{w}}}{67108864 \sqrt{2}}-\frac{60129e^{-3 \Delta w-\Delta  \bar{w}}}{134217728 \sqrt{2}}\cr
&&+\frac{48339 e^{-2 \Delta w-\Delta  \bar{w}}}{67108864 \sqrt{2}}+\frac{154449 e^{-\Delta w-\Delta \bar{w}}}{134217728 \sqrt{2}}+\frac{154449 e^{\Delta w-\Delta  \bar{w}}}{134217728\sqrt{2}}+\frac{48339 e^{2 \Delta w-\Delta  \bar{w}}}{67108864 \sqrt{2}}\cr
&&-\frac{60129 e^{3\Delta w-\Delta  \bar{w}}}{134217728 \sqrt{2}}-\frac{60129 e^{\Delta  \bar{w}-3\Delta w}}{134217728 \sqrt{2}}+\frac{48339 e^{\Delta  \bar{w}-2 \Delta w}}{67108864 \sqrt{2}}+\frac{154449 e^{\Delta  \bar{w}-\Delta w}}{134217728\sqrt{2}}\cr
&&+\frac{154449 e^{\Delta w+\Delta  \bar{w}}}{134217728 \sqrt{2}}+\frac{48339 e^{2\Delta w+\Delta  \bar{w}}}{67108864 \sqrt{2}}-\frac{60129 e^{3 \Delta w+\Delta\bar{w}}}{134217728 \sqrt{2}}-\frac{18819 e^{2 \Delta  \bar{w}-3 \Delta w}}{67108864\sqrt{2}}\cr
&&+\frac{15129 e^{2 \Delta  \bar{w}-2 \Delta w}}{33554432 \sqrt{2}}+\frac{48339 e^{2\Delta  \bar{w}-\Delta w}}{67108864 \sqrt{2}}+\frac{48339 e^{\Delta w+2 \Delta\bar{w}}}{67108864 \sqrt{2}}+\frac{15129 e^{2 \Delta w+2 \Delta  \bar{w}}}{33554432\sqrt{2}}\cr
&&-\frac{18819 e^{3 \Delta w+2 \Delta  \bar{w}}}{67108864 \sqrt{2}}+\frac{23409 e^{3\Delta  \bar{w}-3 \Delta w}}{134217728 \sqrt{2}}-\frac{18819 e^{3 \Delta  \bar{w}-2\Delta w}}{67108864 \sqrt{2}}-\frac{60129 e^{3 \Delta  \bar{w}-\Delta w}}{134217728 \sqrt{2}}\cr
&&-\frac{60129 e^{\Delta w+3 \Delta  \bar{w}}}{134217728\sqrt{2}}-\frac{18819 e^{2 \Delta w+3 \Delta  \bar{w}}}{67108864 \sqrt{2}}+\frac{23409 e^{3\Delta w+3 \Delta  \bar{w}}}{134217728 \sqrt{2}}
\label{aa to aaaa 22 1111}
\eea
Writing this as
\bea
&&\mathcal{A}^{\a\a\to \a \a\a\a}(w_1,w_2,\bar w_1,\bar w_2)\cr
&&\qquad = \sum_{m=-6}^{6} \sum_{m'=-6}^{6}B^{\a\a\to \a\a\a\a}_{m,m'}(n_1=2,n_2=2; p=1,q=1,r=1,s=1)e^{{m  \Delta  w\over2}+{m'\Delta \bar w\over2}}\nn
\label{aa to aaaa condensed}
\eea
defines the coefficients $B^{\a\a\to \a\a\a\a}_{m,m'}(n_1=2,n_2=2; p=1,q=1,r=1,s=1)$.

\subsection{The amplitude for the process $\a\a\to \a\a dd$}

Consider the  amplitude for two bosons going to two bosons and two fermions on the left, and the same split occurring on the right. We take the following initial and final mode numbers
\bea
n_i&=&\lbrace n_1=1,n_2=2\rbrace\cr
n_f &=& \lbrace p=1,q=1,r={1\over2},s={1\over2}\rbrace
\eea
We find \cite{dissertation}
\bea
&&\!\!\!\!\!\!\!\!\!\!\!\!\mathcal{A}^{\a\a\to \a\a dd}(w_1,w_2,\bar w_1,\bar w_2)\cr
&&\!\!\!\!\!\!=\e^{\dot{C}\dot{D}}\e^{\dot{A}\dot{B}}\langle \Phi_4|\big(G^+_{\dot{C},-{1\over2}}\s^-(w_2)G^-_{\dot{A},-{1\over2}}\s^+(w_1)\big)\big(\bar{G}^+_{\dot{D},-{1\over2}}\bar{\s}^-(\bar{w}_2)\bar{G}^-_{\dot{B},-{1\over2}}\bar{\s}^+(\bar{w}_1)\big)|\Psi_2\rangle
\cr
&&\!\!\!\!\!\!=\frac{81}{65536 \sqrt{2}}+\frac{169 e^{-2 \Delta w-2 \Delta  \bar{w}}}{262144 \sqrt{2}}-\frac{19 e^{-\Delta w-2\Delta  \bar{w}}}{65536 \sqrt{2}}+\frac{29 e^{\Delta w-2 \Delta\bar{w}}}{65536 \sqrt{2}}-\frac{119 e^{2 \Delta w-2 \Delta  \bar{w}}}{262144\sqrt{2}}
\cr
&&-\frac{3 e^{-\frac{\Delta w}{2}-\frac{3 \Delta  \bar{w}}{2}}}{4096\sqrt{2}}-\frac{3 e^{\frac{\Delta w}{2}-\frac{3 \Delta  \bar{w}}{2}}}{8192\sqrt{2}}-\frac{19 e^{-2 \Delta w-\Delta  \bar{w}}}{65536 \sqrt{2}}+\frac{5 e^{-\Delta w-\Delta\bar{w}}}{16384 \sqrt{2}}-\frac{3 e^{\Delta w-\Delta  \bar{w}}}{16384\sqrt{2}}+\frac{29 e^{2 \Delta w-\Delta  \bar{w}}}{65536 \sqrt{2}}
\cr
&&-\frac{3 e^{-\frac{3 \Delta w}{2}-\frac{\Delta\bar{w}}{2}}}{4096 \sqrt{2}}+\frac{5 e^{-\frac{\Delta w}{2}-\frac{\Delta \bar{w}}{2}}}{8192 \sqrt{2}}+\frac{e^{\frac{\Delta w}{2}-\frac{\Delta  \bar{w}}{2}}}{2048 \sqrt{2}}-\frac{3 e^{\frac{3 \Delta w}{2}-\frac{\Delta  \bar{w}}{2}}}{8192 \sqrt{2}}-\frac{3 e^{-\frac{3\Delta w}{2}+\frac{\Delta  \bar{w}}{2}}}{8192 \sqrt{2}}+\frac{e^{-\frac{\Delta w}{2}+\frac{\Delta  \bar{w}}{2}}}{2048 \sqrt{2}}\cr
&&+\frac{5 e^{\frac{\Delta w}{2}+\frac{\Delta  \bar{w}}{2}}}{8192\sqrt{2}}-\frac{3 e^{\frac{3 \Delta w}{2}+\frac{\Delta  \bar{w}}{2}}}{4096\sqrt{2}}+\frac{29 e^{-2 \Delta w+\Delta  \bar{w}}}{65536 \sqrt{2}}-\frac{3 e^{-\Delta w+\Delta \bar{w}}}{16384 \sqrt{2}}+\frac{5 e^{\Delta w+\Delta  \bar{w}}}{16384\sqrt{2}}-\frac{19 e^{2 \Delta w+\Delta  \bar{w}}}{65536 \sqrt{2}}
\cr
&&-\frac{3 e^{-\frac{\Delta w}{2}+\frac{3\Delta  \bar{w}}{2}}}{8192 \sqrt{2}}-\frac{3 e^{+\frac{\Delta w}{2}+\frac{3 \Delta \bar{w}}{2}}}{4096 \sqrt{2}}+\frac{9 e^{\frac{3 \Delta w}{2}+\frac{3 \Delta \bar{w}}{2}}}{8192 \sqrt{2}}-\frac{119 e^{-2\Delta w+2 \Delta  \bar{w}}}{262144 \sqrt{2}}+\frac{29 e^{-\Delta w+2 \Delta  \bar{w}}}{65536\sqrt{2}}
\cr
&&-\frac{19 e^{\Delta w+2 \Delta  \bar{w}}}{65536 \sqrt{2}}+\frac{169 e^{2 \Delta w+2 \Delta \bar{w}}}{262144 \sqrt{2}}-\frac{45 e^{-2 \Delta  \bar{w}}}{131072\sqrt{2}}-\frac{9 e^{-\Delta  \bar{w}}}{32768\sqrt{2}}-\frac{9 e^{\Delta  \bar{w}}}{32768\sqrt{2}}-\frac{45 e^{2 \Delta  \bar{w}}}{131072 \sqrt{2}}\cr
&&-\frac{45 e^{-2 \Delta w}}{131072\sqrt{2}}-\frac{9 e^{-\Delta w}}{32768 \sqrt{2}}-\frac{9 e^{\Delta w}}{32768\sqrt{2}}-\frac{45 e^{2 \Delta w}}{131072 \sqrt{2}}+\frac{9 e^{-\frac{3 \Delta w}{2}-\frac{3 \Delta  \bar{w}}{2}}}{8192\sqrt{2}}
\label{aa to aadd 12 11halfhalf}
\eea
Writing this as 
\bea
&&\mathcal{A}^{\a\a\to \a\a dd}(w_1,w_2,\bar w_1,\bar w_2)\cr
&&\qquad = \sum_{m=-4}^{4} \sum_{m'=-4}^{4}B^{\a\a\to \a\a dd}_{m,m'}(n_1=1,n_2=2; p=1,q=1,r={1\over2},s={1\over2})e^{{m  \Delta  w\over2}+{m'\Delta \bar w\over2}}\nn
\label{aa to aadd condensed}
\eea
defines the coefficients $B^{\a\a\to \a\a dd}_{m,m'}(n_1=1,n_2=2; p=1,q=1,r={1\over2},s={1\over2})$.

\section{ Integrating the amplitude}\label{integration}

We now perform the integration over the positions of the deformation operators $D$.  We note that in the steps below we wick rotate $\t$ back to our physical time coordinate $t$ through the transformation
\bea
\t\to it
\eea
Our full integrated amplitude will have the form
\bea
\mathcal{A}^{i\to j}_{\text{int}} &=&{1\over2}\lambda^2\int_{-{\t\over2}}^{{\t\over2}}\int_{-{\t\over2}}^{\t_2}\int_{\s_2=0}^{2\pi}\int_{\s_1=0}^{2\pi}d^2w_2d^2w_1\mathcal{A}^{i\to f}(w_1,w_2,\bar w_1,\bar w_2) \cr
&=&{1\over2}\lambda^2\sum_{m=m_{min}(n_{i};n_{f})}^{m_{max}(n_{i};n_{f})}\sum_{m'=m'_{min}(n_{i};n_{f})}^{m'_{max}(n_{i};n_{f})}B^{i\to f}_{m,m'}(n_i;n_f)\cr
&&\quad \int_{-{\t\over2}}^{{\t\over2}}\int_{-{\t\over2}}^{\t_2}\int_{\s_2=0}^{2\pi}\int_{\s_1=0}^{2\pi}d^2w_2d^2w_1 e^{{m\Delta w\over2}} e^{{m'\Delta \bar{w}\over2}}\cr
&\equiv&{1\over2}\lambda^2\sum_{m=m_{min}(n_{i};n_{f})}^{m_{max}(n_{i};n_{f})}\sum_{m'=m'_{min}(n_{i};n_{f})}^{m'_{max}(n_{i};n_{f})}B^{i\to f}_{m,m'}(n_i;n_f) I_{m,m'}(t)
\label{int amp}
\eea
where

\bea
I_{m,m'}(t)&=&{1\over R^2}\int_{-{t\over2}}^{{t\over2}}dt_2\int_{-{t\over2}}^{t_2}dt_1\int_{\s_2=0}^{2\pi}\diff \s_2\int_{\s_1=0}^{2\pi}\diff \s_1 e^{{i(m+m')\over2}{t'\over R}}e^{{i(m-m')\over2}\s'}
\label{integral}
\eea
with $m,m'$ taking integer values. We also note that $t'=t_2 - t_1$ and $\s'=\s_2 -\s_1$. 


\subsection{Evaluating $I_{m,m'}$}\label{evaluatingI}
When evaluating the integral (\ref{integral}), we have four cases:
\subsubsection{\underline{$m-m'\neq 0,\quad m+m' \neq 0$}}
\bea
I_{m-m'\neq 0,m+m' \neq 0}(t)&=&{1\over }\int_{-{t\over2}}^{{t\over2}}dt_2\int_{-{t\over2}}^{t_2}dt_1\int_{\s=0}^{2\pi}d\s_2\int_{\s=0}^{2\pi}d\s_1 e^{{i(m+m')\over2}{t'\over }}e^{{i(m-m')\over2}\s'}\cr
&=&{32i\sin^2\big({(m-m')\over2}\pi\big)\bigg((m+m')t - 4e^{i{(m+m')t\over 4}}\sin\big({(m+m')t\over 4}\big)\bigg)\over (m^2 - m'^2)^2}\nn
\label{I1}
\eea

\subsubsection{\underline{$m-m'\neq 0, m+m' = 0$}}

\bea
I_{m-m'\neq 0, m+m' = 0}(t)&=&\int_{-{t\over2}}^{{t\over2}}dt_2\int_{-{t\over2}}^{t_2}dt_1\int_{\s=0}^{2\pi}d\s_2\int_{\s=0}^{2\pi}d\s_1e^{{i(m-m')\over2}\s'}\d_{m+m',0}\cr
&&=-{t^2(\cos(2m\pi)-1)\over m^2}\cr
&&=0
\label{integral two}
\eea

\subsubsection{\underline{$m-m'=0 , m+m' \neq0$}}
\bea
I_{m-m'=0 , m+m' \neq0}(t)&=&\int_{-{t\over2}}^{{t\over2}}dt_2\int_{-{t\over2}}^{t_2}dt_1\int_{\s=0}^{2\pi}d\s_2\int_{\s=0}^{2\pi}d\s_1 e^{{i(m+m')\over2}t'}\d_{m-m',0}\cr
&=& {4i \pi^2\bigg(m{t\over }  -  2e^{i {mt\over2}}\sin(  {mt\over 2})  \bigg)\over m ^2}
\label{integral three}
\eea

\subsubsection{\underline{$m=m'=0$}}

\bea
I_{m=m'=0}(t)&=&{1\over }\int_{-{t\over2}}^{{t\over2}}dt_2\int_{-{t\over2}}^{t_2}dt_1\int_{\s=0}^{2\pi}d\s_2\int_{\s=0}^{2\pi}d\s_1\cr
&=& 2\pi^2 t^2
\label{integral four}
\eea

\subsection{Energy of the intermediate state}\label{intermediate}

The integrals $I_{m,m'}$ obtained above are all that we need to complete our evaluation of amplitudes. But before proceeding it is useful to note the physical origin of the oscillating and the growing terms in these integrals. Let initial states be placed at $t_i=-{t\over2}$ and final states be placed at $t_f = {t\over2}$. Let the initial and final states have energy $E$. Between the two deformation operators, i.e., in the interval $t_1<t'<t_2$,  we have the propagation of an intermediate state whose energy we call $E'$. We will see that the states with $E'\ne E$ generate oscillatory terms and terms proportional to $t$, while states with $E'=E$ yield a term proportional to $t^2$.

The evolution in the interval $-{t\over 2}<t'<t_1$ gives a factor
\bea
e^{-iE(t_1+{t\over2})}
\eea
The evolution in the interval $t_1<t'<t_2$ gives a factor
\bea
e^{-iE'(t_2 - t_1)}
\eea
Finally, the evolution in the interval $t_2<t'<{t\over 2}$ gives a factor
\bea
e^{-iE({t\over2}-t_2) }
\eea
Thus the amplitude for an intermediate state with energy $E'$ is, apart from an overall constant
\bea
&&e^{-iE(t_1+{t\over2})}e^{-iE'(t_2 - t_1)}e^{-iE({t\over2}-t_2) }=e^{i(E'-E)t_1}e^{i(E-E')t_2}e^{-iEt}
\label{evolution factors}
\eea
Now we consider two cases:

\b

(A) { $E\neq E'$:}  We first integrate (\ref{evolution factors}) over $t_1$ from $[-{t\over2} ,t_2]$. This gives
\bea
&&e^{i(E-E')t_2}e^{-iEt}\int_{-{t\over2}}^{t_2}dt_1e^{i(E'-E)t_1}\cr
&&=e^{i(E-E')t_2}e^{-iEt}{1\over i(E'-E)}\bigg(e^{i(E'-E)t_2} - e^{-i(E'-E){t\over 2}}\bigg)\cr
&&=e^{-iEt}{1\over i(E'-E)} -e^{-iEt}{1\over i(E'-E)}e^{i(E-E')t_2}e^{-i(E'-E){t\over 2}}
\eea
Integrating over $t_2$ from $[-{t\over2} ,{t\over2}] $ yields
\bea
&&e^{-iEt}{1\over i(E'-E)}\int_{-{t\over2}}^{{t\over2}}dt_2 ~~-~~ e^{-iEt}{1\over i(E'-E)}e^{-i(E'-E){t\over 2}}\int_{-{t\over2}}^{{t\over2}}dt_2e^{i(E-E')t_2}\cr
&&e^{-iEt}\bigg({1\over i(E'-E)}t ~~-~~{1\over (E'-E)^2}e^{-i(E'-E){t\over 2}}2i\sin[(E-E'){t\over2}]\bigg)
\eea
Thus we get a term  which is proportional to $t$ and another term which oscillates in $t$; we do not however get a term that grows as $t^2$. 

\b

(B) { $E = E'$:}  Again, integrating (\ref{evolution factors}) over $t_1$ from $[-{t\over2} ,t_2]$ yields
\bea
&&e^{-iEt}\int_{-{t\over2}}^{t_2}dt_1\cr
&&\quad = e^{-iEt}(t_2 + {t\over2})
\eea
Integrating over $t_2$ from $[-{t\over2} ,{t\over2}] $ yields
\bea
 &&e^{-iEt}\int_{-{t\over2}}^{{t\over2}}dt_2(t_2 + {t\over2})\cr
 && e^{-iEt}{t^2\over2}
\eea
Thus we get a term that grows as $t^2$ but we do not get oscillating terms or terms that grow as $t$.


\subsection{Integrated Amplitude for $\a\to\a\a\a$}

In this subsection we compute the the full integrated amplitude for the process where (both on the left and right sides) {one} boson with energy $n=3$ splits into {three} bosons, each with energy $p=q=r=1$. We need to evaluate the expression (\ref{a to aaa 3 to 111}). 

In (\ref{a to aaa 3 to 111}) we note that  $m,m'\in \mathbb{Z}_{\text{odd}}$, which implies that $m-m'\in \mathbb{Z}_{\text{even}}$. This gives
\bea
I_{m-m'\neq 0,m+m' \neq 0}(t) &=& {32i\sin^2\big({(m-m')\over2}\pi\big)\bigg((m+m')t - 4e^{i{(m+m')t\over 4}}\sin\big({(m+m')t\over 4}\big)\bigg)\over (m^2 - m'^2)^2}
\cr
\cr
&=&0,\quad m,m'\in\mathbb{Z}_{\text{odd}},\quad m-m'\neq 0 \neq m+m'
\eea
Note that the terms with $m+m'=0$ vanish according to (\ref{integral two}). Also, since there are no $m=m'=0$ terms in (\ref{a to aaa 3 to 111}), we have no contribution from (\ref{integral four}). Thus the only contributions come from terms of the type (\ref{integral three}). Thus the amplitude (\ref{int amp}) becomes
\bea
\mathcal{A}^{\a\to\a\a\a }_{\text{int}} &=& 8\lambda^2\pi^2  \bigg({39\over 32768}\sin^2\bigg(  {t\over 2}\bigg)  + \frac{159}{131072}{\sin^2(  {3t\over 2})  \over 9} + {3\over131072}\sin^2\bigg(  {5t\over 2}\bigg)  \bigg)
\eea


We see that the amplitude only contains terms which oscillate with time, $t$. We will discuss the implications of this in the next section when we compare this result with that for processes where two excitations split into four excitations.


\subsection{Integrated Amplitude for $\a\to\a dd$}

For the process $\a\to \a dd$ we obtain
\bea
\mathcal{A}^{\a\to \a dd}_{\text{int}} 
&=& 8\lambda^2  \pi^2\bigg( {5\over 4096}\sin^2\bigg(  {t\over 2}\bigg) +   {1\over 4096}\sin^2\bigg(  {3t\over 2}\bigg)  \bigg)
\eea
We see  behavior similar to that of the  process for $\a\to\a\a\a$. The amplitude  oscillates with $t$.

\subsection{Integrated Amplitude for $\a\a\to\a\a \a\a$}

For the process $\a\a\to \a\a \a\a$, we obtain
\bea
\mathcal{A}^{\a\a\to \a \a\a\a}_{\text{int}} &=& \lambda^2 \pi^2t^2\frac{59049}{8388608 \sqrt{2}} 
\cr
&&+ 8  \lambda^2 \pi^2\Bigg( \frac{154449}{134217728 \sqrt{2}}{ \sin^2(  t)   \over 4} + \frac{15129}{33554432\sqrt{2}}{ \sin^2(  2t) \over 16} +  \frac{23409}{134217728 \sqrt{2}}{\sin^2(  3t)   \over 36}\Bigg)\nn
\eea

In this case we see that the amplitude  includes a $t^2$ term in addition to terms which oscillate in $t$. 


\subsection{Integrated Amplitude for $\a\a\to\a\a dd$}

For the process $\a\a\to \a\a dd$, we obtain
\bea
\mathcal{A}^{\a\a\to \a\a dd}_{\text{int}} 
&=&\lambda^2\pi^2\frac{81}{65536 \sqrt{2}}t^2 \cr
&& + 8 \pi^2\lambda^2 \bigg(\frac{5}{8192 \sqrt{2}}   \sin^2\bigg(  {t\over 2}\bigg)    + \frac{5}{16384 \sqrt{2}} { \sin^2(t)   \over 4}  \cr
&&\qquad\qquad+ \frac{1 }{8192\sqrt{2}} \sin^2\bigg(  {3t\over 2}\bigg)    + \frac{169}{262144 \sqrt{2}}  {  \sin^2(  2t)  \over 16}   \bigg)\nn
\eea
As in the case of the process  $\a\a\to\a\a\a\a$, we find that this amplitude has terms which oscillate in $t$ and a term which grows like $t^2$. 

\section{Coefficients for Larger Energies for $\a\a\to\a\a\a\a$}

In the above sections we looked at low energies of excitation. This enabled us to list explicit results for the amplitudes. The amplitudes are computed with the help of a symbolic manipulation package, so it is straightforward to compute them for higher energies of excitation. We now summarize some features of the amplitudes as we raise the energy.  Some interesting patterns can be observed from the growth of these amplitudes with energy. 

We have seen that the amplitudes which correspond to two particles in the initial state have a $t^2$ secular growth.  We tabulate the coefficients of  $t^2$  for bosonic splitting amplitudes in the $2\to4$ process, $B^{\a\a\to \a\a\a\a}_{0,0}\big(n_1,n_2; p,q,r,s\big)$, for increasing values of total initial energy $E_{\text{total}}$. 

 We compare these amplitudes for different modes of splitting. We have several excitations in the final state. Is there a higher amplitude for distributing the energies roughly equally among these final state excitations, or is there a higher amplitude when the energy is predominantly carried by one excitation? 
 
  We tabulate our results below.

\begin{table}[H]
\centering
 \begin{tabular}{||c   | c||} 
 \hline
 $E_{\text{total}} = 2(n_1+n_2) $ & \underline{Equal Splitting}\\ 
  $( n_1 = n_2)$ & $B^{\a\a\to \a\a\a\a}_{0,0}\big(n_1,n_2; p=q=r=s={n_1 + n_2\over4}\big)$ \\[0.5ex] 
 \hline\hline
 $2(n_1+n_2)=2(2+2)=8$ & $4.97746\times 10^{-3}$  \\ 
 \hline
 $2(n_1+n_2)=2(6+6)=24$ &  $7.59583\times 10^{-5}$\\
 \hline
 $2(n_1+n_2)=2(10+10)=40$ & $1.15986\times 10^{-5}$  \\
 \hline
$2(n_1+n_2)=2(14+14)=56$ & $3.36791\times10^{-6}$  \\
 \hline
$2(n_1+n_2)=2(18+18)=72$ & $1.33531\times10^{-6}$  \\ [1ex] 
 \hline
\end{tabular}
\caption{\label{equalsplitting} $t^2$ coefficients for $2\to 4$ splitting for equal energies in the final state: $ p=q=r=s={n_1 + n_2\over4}$. The factor of $2$ in the total energy comes because we must add together the left and right moving energies which are chosen to be the same.}
\end{table}

\begin{table}[H]
\centering
 \begin{tabular}{||c   | c||} 
 \hline
$E_{\text{total}} = 2(n_1+n_2) $ & \underline{Asymmetric Splitting}\\ 
  $( n_1 = n_2)$ & $B^{\a\a\to \a\a\a\a}_{0,0}\big(n_1,n_2; p=n_1 + n_2-3,q=r=s=1\big)$ \\[0.5ex] 
 \hline\hline
 $2(n_1+n_2)=2(2+2)=8$ & $4.97746\times 10^{-3}$\\ 
 \hline
 $2(n_1+n_2)=2(6+6)=24$ & $1.28367\times10^{-8}$  \\
 \hline
 $2(n_1+n_2)=2(10+10)=40$ & $2.33709\times10^{-10}$  \\
 \hline
$2(n_1+n_2)=2(14+14)=56$ & $1.86499\times10^{-11}$  \\
 \hline
$2(n_1+n_2)=2(18+18)=72$ & $2.92488\times10^{-12}$  \\ [1ex] 
 \hline
\end{tabular}
\caption{\label{asymmetricsplitting} $t^2$ coefficients for $2\to 4$ splitting for unequal energies in the final state: $ p=n_1 + n_2-3, q=r=s=1$. Again, the factor of $2$ in the total energy comes because we must add together the left and right moving energies which are chosen to be the same.}
\end{table}

We see that at each energy level, $E_{\text{total}}$, the $t^2$ coefficient for \textit{equal} splitting is much greater than the $t^2$ coefficient for \textit{asymmetric} splitting. This tells us that the probability for the total energy of two initial modes to split equally amongst four final modes is much higher than to split asymmetrically.\footnote{The  lowest level $E_{\rm total}=8$ has only one way for the final state to split, so we see only the case with  equal coefficients.} 

This nature of splitting serves as an  indicator of the way thermalization is expected to progress, as we will discuss in the next section.

\section{Discussion}

In this paper we addressed the problem of black hole formation in AdS by investigating thermalization in the dual CFT. To understand thermalization we considered the twist deformation, the fundamental interaction of the theory.
We looked at two main scattering processes at \textit{second} order in the twist deformation: 1) one excitation splitting into three excitations and 2) two excitations splitting into four excitations. The `1 to 3' processes were found to only have terms which oscillate in $t$. The `2 to 4' processes were found to have similar oscillatory terms. In addition however, they were  found to also have `secular' terms proportional to $t^2$. These $t^2$ terms arise from the existence of accessible intermediate states which have the same energy as the initial and final states. 

Let us now put these results into a picture of thermalization. First consider thermalization in  a gas of atoms. Suppose a particle with high energy $E$ enters a gas which is at a low temperature $T$; thus the incoming particle has much higher velocity $v$ than the typical velocity $\bar v$ of the particles in the gas.  When the particle with velocity $v$ collides with one of the slow particles, their dynamics is given by an interaction vertex; generically, we find that the two scattered particles have velocities $\sim v$ each (rather than an asymmetric situation with one velocity small and one large). This basic scattering vertex allows us to get a qualitative picture of thermalization. The vertex involves computing the scattering probability for any choice of final velocities $v_1, v_2$ and angles $\theta_1, \phi_1, \theta_2, \phi_2$. Once we have computed this vertex, we can concatenate several such vertices to get a picture of thermalization: at each step the energy in a high energy particle splits among the energies of the scattered particles. 

In this language, what we have done in the present paper is compute the scattering vertex for the CFT with $n_1n_5=N=2$. We hope to present the scattering matrix for higher $N$ in later works. These results should allow us to get a qualitative picture of how thermalization proceeds in the CFT. 

A key aspect we have noted in the introduction is that we have to be careful to identify the effects that correspond to genuine thermalization. Infall in the metric (\ref{one}) gives a redshift going to infinity, but the infalling object remains intact; we have argued that this process of increasing redshift should not be called thermalization. But many computations in the CFT that aim to see black hole formation focus just on getting a situation with large redshift. We have argued, on the other hand, that thermalization in the CFT will start after the step in gravity where we get a large redshift. One will have to see effects of finite $N$, which we enforced by taking a low value $N=2$. The scattering vertex we computed gives a very rough qualitative guide to thermalization, but this is a picture we hope to improve upon in later works. 

\section{Acknowledgements}
We would like to thank Zaq Carson, David Turton and Bin Guo for may helpful discussions. The work of S.D.M is supported by DOE grant de-sc0011726. The work of S.H is supported by the Presidential Fellowship and the Lyman T. Johnson Postdoctoral Fellowship.


\end{document}